\documentclass[12pt]{iopart}

\usepackage{epsfig,setstack,iopams}
\usepackage{subfigure}

\jl{6}   % Submit to CQG
 
\newcommand{\deriv}[2]{\frac{d#1}{d#2}}
\newcommand{\pderiv}[2]{\frac{\partial #1}{\partial #2}}
\begin{document}
\title[]{A cosmological solution of Regge calculus}
\author{Adrian P Gentle}
\address{Department of Mathematics, University of Southern Indiana,
    Evansville, IN 47712}
  \ead{apgentle@usi.edu}
\date{\today}
\begin{abstract} 
  We revisit the Regge calculus model of the Kasner cosmology first
  considered by S. Lewis.  One of the most highly symmetric
  applications of lattice gravity in the literature, Lewis' discrete
  model closely matched the degrees of freedom of the Kasner
  cosmology.  As such, it was surprising that Lewis was unable to
  obtain the full set of Kasner-Einstein equations in the continuum
  limit. Indeed, an averaging procedure was required to ensure that
  the lattice equations were even consistent with the exact solution
  in this limit. We correct Lewis' calculations and show that the
  resulting Regge model converges quickly to the full set of
  Kasner-Einstein equations in the limit of very fine discretization.
  Numerical solutions to the discrete and continuous-time lattice
  equations are also considered.
\end{abstract}
%\submitto{\CQG} 
%\setcounter{footnote}{0}
%\maketitle
%%%%%%%%%%%%%%%%%%%%%%%%%%%%%%%%%%%%%%%%%%

\section{Introduction} 

The discrete formulation of gravity proposed by T.~Regge in 1961
\cite{regge61} has been deployed in a wide variety of settings, from
probing the foundations of gravity and the quantum realm
\cite{williams92,regge00} to numerical studies of classical
gravitating systems \cite{williams92,gentle02}.  Regge calculus continues to be
used in new and diverse ways; recent examples include Ricci flow
\cite{miller11} and as an explanation of dark energy
\cite{stuckey11}.

In this paper we re-examine the Regge calculus model of the vacuum
Kasner cosmology first considered by Lewis \cite{lewis82}, with the
goal of gaining insight into the continuum limit of this discrete
approach to gravity.  In a general setting the structural differences
between a continuous manifold and a discrete simplicial lattice lead
to difficulties in directly comparing the Regge and Einstein equations
or their solutions, with a single Regge equation per edge in the
lattice compared with ten Einstein equations per event in spacetime.
We expect many more simplicial equations than Einstein equations in a
general simulation, and some form of averaging must be expected before
the equations (or their solutions) can be compared.

Lewis \cite{lewis82} studied both the Kasner and spatially flat
Friedmann-Lema\^itre-Robertson-Walker (FLRW) cosmologies using a
regular hypercubic lattice. We only consider the Kasner solution in
this paper, where the high degree of symmetry, without the added
complication of matter, allows explicit examination of the Regge
equations in the continuum limit.  By aligning the degrees of freedom
of the lattice with the continuum metric components, Lewis was able to
avoid the issue of averaging and make direct comparisons between the
Regge equations and the Kasner-Einstein equations in the continuum
limit.  Unfortunately, Lewis was only able to recover one of the four
Einstein equations in this limit, and even this was only possible
after the equations were carefully averaged \cite{lewis82}.  Without
this averaging it is not clear that the equations obtained by Lewis
actually represent the Kasner cosmology.

We show that Lewis neglected a vital portion of the simplicial
curvature arising from the two-dimensional spacelike lattice faces
that lie on constant time hypersurfaces.  The Kasner cosmology has
zero intrinsic curvature on constant time hypersurfaces, so the
lattice curvature concentrated on spacelike faces -- measured on a
plane with signature ``-+'' orthogonal to the face -- make an
important contribution to the total lattice curvature.  We show that
when these curvature terms are included, the discrete equations
exactly reproduce the Kasner-Einstein equations in the limit of very
fine triangulations without the need to average.

In addition to reconsidering Lewis' analytic work on the spatially
flat, anisotropic Kasner cosmology \cite{lewis82}, we construct
numerical solutions to the discrete and continuous-time lattice
equations.  This builds on the previous work of Collins and Williams
\cite{collins73} and Brewin \cite{brewin87} on highly symmetric,
continuous time, closed FLRW cosmologies, and the (3+1)-dimensional
numerical study of the Kasner cosmology by Gentle \cite{gentle98} with
discrete time and coarse spatial resolution.  We begin by briefly
describing the continuum Kasner solution.

\section{The Kasner cosmology}
\label{sec:kasner}

The Kasner solution \cite{kasner21} is a vacuum, homogeneous,
anisotropic cosmological solution of the Einstein equations with
topology $R\times T^3$.  Appropriate slicing creates flat spacelike
hypersurfaces, while the global topology allows non-trivial vacuum
solutions of the Einstein equations.

The Kasner metric may be written in the form \cite{mtw}
\[
  ds^2 = - dt^2 + f(t)^2 \,dx^2 + g(t)^2\, dy^2 + h(t)^2\, dz^2
\]
where the functions $f$, $g$ and $h$ are determined by the vacuum Einstein
equations
\begin{eqnarray}
  0 & =\ G_{tt} 
  = & f \deriv{g}{t} \deriv{h}{t} + g \deriv{f}{t} \deriv{h}{t}
  + h \deriv{f}{t} \deriv{g}{t}, \label{eqn:kasnert}  \\
  0 & =\ G_{xx} 
  = & \deriv{g}{t} \deriv{h}{t} + h \deriv{^2g}{t^2} 
  + g \deriv{^2h}{t^2},    \label{eqn:kasnerx}  \\
  0 & =\ G_{yy} 
  = & \deriv{f}{t} \deriv{h}{t} + h \deriv{^2f}{t^2} 
  + f \deriv{^2h}{t^2},    \label{eqn:kasnery} \\
  0 & =\ G_{zz}
  = & \deriv{f}{t} \deriv{g}{t} + f \deriv{^2g}{t^2} 
  + g \deriv{^2f}{t^2}.   \label{eqn:kasnerz} 
\end{eqnarray}
Note that the equation $G_{tt}=0$ is a first integral of the remaining
equations.  The Kasner metric components are 
\[
f(t) = t^{2p_1}, \qquad g(t) = t^{2p_2}, \qquad h(t) = t^{2p_3},
\nonumber
\]
where the Kasner exponents $p_i$ are unknown constants.  With this
choice of metric functions the vacuum Kasner-Einstein equations reduce
to two algebraic constraints,
\[
p_1^2 + p_2^2 + p_3^2 = p_1 + p_2 + p_3 = 1,
\]
leaving a one parameter family of Kasner cosmologies.  

The Kasner solutions are the basis of the Mixmaster cosmologies
\cite{belinsky70}, which may be regarded as a series of Kasner-like
epochs undergoing an infinite series of ``bounces'' from one set of
Kasner exponents to the next.  It is conjectured that these asymptotic
velocity term dominated models embody the generic approach to
singularity in crunch cosmologies, and it has been shown that the
bounces represent a chaotic map on the Kasner exponents
\cite{cornish97,berger02}.

%%%%%%%%%%%%%%%%%%%%%%%%%%%%%%%%%%%%%%%%%%%%%%%%%%%%%%%

\section{A homogeneous, anisotropic spacetime lattice}
\label{sec:lattice}

We follow Lewis and build a discrete approximation of the Kasner
spacetime using a highly symmetric lattice of rectangular prisms.  The
regularity of the lattice implements homogeneity, while the
rectangular prisms allow a degree of anisotropy.  The complete
four-geometry is constructed by extruding the initial three-geometry
forward in time and filling the interior with four-dimensional prisms.

Each flat $T^3$ hypersurface consists of a single rectangular prism
with volume $x_iy_iz_i$, where opposing faces are identified to give
the global topology.  This prism is subdivided into $n^3$ regular
prisms with edge lengths $u_i, v_i$ and $w_i$, with
\begin{equation}
  \label{eqn:ratios}
  u_i = \frac{x_i}{n}, \qquad   v_i = \frac{y_i}{n}, \qquad
  w_i = \frac{z_i}{n},
\end{equation}
and where the subscript $i$ labels the Cauchy surface at time
$t_i$. 

\begin{figure}[t]
  \begin{center}
    \includegraphics[width=3in]{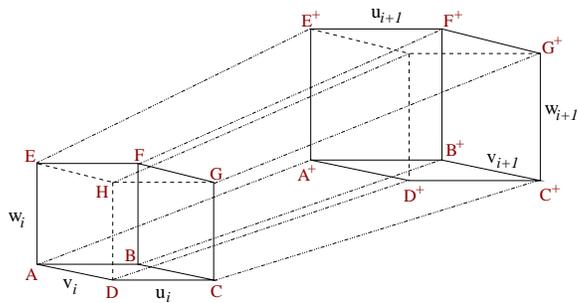}
  \end{center}
  \caption{The section of a world-tube joining a rectangular prism to
    its future counterpart.  Homogeneity implies that an observer will
    fall freely along the centre of the worldtube, providing a
    convenient coordinate system from which to view the lattice.}
  \label{fig:block}
\end{figure}

The three-geometry is joined to a similar structure at time
$t=t_{i+1}= t_i + \Delta t_i$, where the prisms have edge lengths
$u_{i+1}$, $v_{i+1}$ and $w_{i+1}$.  This structure is shown in figure
\ref{fig:block}.  Time evolution of the initial surface maintains
homogeneity, so within the worldtube of each prism there exists a
local freely-falling inertial frame.  In the coordinates of this frame
the coordinates of vertex $A$ in figure \ref{fig:block} can be written
as
\[
  \left( t_i, \frac{u_i}{2}, -\frac{v_i}{2}, -\frac{w_i}{2}
  \right) 
\]
and similarly, the coordinates for vertex $A^+$ (the counterpart of
$A$ on the next hypersurface) are
\[
  \left( t_{i+1}, \frac{u_{i+1}}{2}, 
    -\frac{v_{i+1}}{2}, -\frac{w_{i+1}}{2} \right).
\]
The spacetime interval along the time-like edge joining $A$ and $A^+$
is defined to be $m_i^2<0$, and thus
\begin{equation}
  \label{eqn:timelike_edge}
  m_i^2 = - \Delta t_i^2 + \frac{1}{4} \left( \Delta u_i ^2 +
    \Delta v_i ^2 + \Delta w_i ^2 \right),
\end{equation}
where the difference operator is defined as $\Delta l_i = l_{i+1} -
l_i$.  Note that the requirement that $m_i^2 < 0$ implies a
restriction on the choice of $\Delta t_i$ for a given value of the
``resolution'' parameter $n$.  Homogeneity guarantees that identical
expressions hold for all timelike edges joining the spacelike
hypersurfaces labeled $t_i$ and $t_{i+1}$.

The discrete spacetime curvature in Regge calculus is manifest on the
two-dimensional faces on which three-dimensional blocks hinge
\cite{regge61}, and is represented by the angle deficit (the
difference from the flat space value) measured in the plane orthogonal
to the face.  There are two distinct classes of two-dimensional faces,
or hinges, in the lattice: timelike areas formed by evolving a
spacelike edge forward to the next hypersurface, and the rectangular
faces of the prisms on $t=\mbox{constant}$ slices.

The timelike trapezoids formed when the spatial edge $u_i$ is carried
forward in time from the hypersurface labeled $t_i$ to $t_{i+1}$,
face $ABA^+B^+$ in figure \ref{fig:block}, has area
\[
  A^{xt}_i = \frac{1}{4} \left( u_{i+1} + u_i \right)
  \sqrt{\Delta u_i^2 - 4m_i^2 },
\]
which is real since $m_j^2<0$.  Likewise, the spacelike hinge $ABCD$
shown in figure \ref{fig:block}, consisting of the edges $u_i$ and
$v_i$, has area
\[
  A^{xy}_i  =  u_i v_i,
\]
with the other spacelike and timelike hinges defined similarly.

%%%%%%%%%%%%%%%%%%%%%%%%

Turning now to the curvature about these faces, we note that there are
four distinct three-dimensional prisms which hinge on the timelike
face $ABA^+B^+$.  Each of these is formed by dragging one of the
two-dimensional faces containing the edge $AB$ forward in time. This
includes the prisms $ABCDA^+B^+C^+D^+$ and $ABEFA^+B^+E^+F^+$.  The
remaining two prisms hinging on $ABA^+B^+$ are not shown in figure
\ref{fig:block}.

The homogeneity of the lattice ensures that the four hyper-dihedral
angles that surround the hinge $ABA^+B^+$ are the same.  Denoting each
of these angles as $\theta^{xt}_i$, the angle defect (or deficit
angle) about the timelike face $ABA^+B^+$ is
\[
  \epsilon^{xt}_i = 2\pi - 4 \theta^{xt} _i,
\]
which measures the deviation of the total angle from the flat-space
value of $2\pi$.  The hyper-dihedral angle $\theta^{xt} _i$ is
\[
  \cos \theta ^{xt}_i = 
  \frac{\Delta v_i \Delta w_i} { 
    \sqrt{ \left( 4\Delta t_i^2 - \Delta v_i^2 \right) 
      \left( 4\Delta t_i^2 - \Delta w_i^2 \right) } 
    },
\]
with analogous definitions for $\theta^{yt}_i$ and $\theta^{zt}_i$,
the hyper-dihedral angles about the remaining classes of spacelike
hinge.

To measure the deficit angles about the spacelike faces, consider the
curvature $\epsilon^{xy}_i$ about the hinge $ABCD$.  Since this hinge
is spacelike, the hyper-dihedral angles are boosts in the plane with
signature $-+$ orthogonal to the hinge.  The deficit angle for a
spacelike hinge is \cite{brewin11}
\[
\epsilon^{xy}_i = - \sum _k \phi_{k},
\]
where the summation is over all boosts $\phi_{k}$ which
surround the hinge.  The boost between the two three-dimensional
prisms which hinge on $ABCD$, namely $ABCDA^+B^+C^+D^+$ and
$ABCDEFGH$, is 
\[
\sinh \phi^{xy} _i = - \frac { \Delta w_i } { \sqrt{4\Delta t_i^2 - 
    \Delta w_i^2} }, 
\]
and the hinge $ABCD$ is surrounded by four such boosts, two identical
boosts above and two below.  Thus the final deficit angle measured
about $ABCD$ is
\[
  \epsilon^{xy}_i =  2 \left ( \phi^{xy} _{i-1} - \phi^{xy} _i
  \right).
\]

Further details on calculating the deficit angles about a spacelike
hinge are contained in a recent paper by Brewin \cite{brewin11}.  We
note that this type of hinge was not included in the calculations of
Lewis \cite{lewis82}, leading to errors in the resulting Regge
equations.  We return to this issue below.

%%%%%%%%%%%%%%%%%%%%%%%%%%%%%%%%%%%%%%%%%%%%%%%%%%%%

\section{The Regge calculus model}
\label{sec:regge}

The vacuum Regge equations take the form \cite{regge61}
\[ 
0 = \sum_t \epsilon_t \, \frac{\partial A_t}{\partial L^2_j},
\]
where the sum is over all triangles $t$ with area $A_t$ that contain
the edge $L_j$, and $\epsilon_t$ the deficit angle about triangle $t$.
Each edge in the lattice yields a single Regge equation, but the
homogeneity and anisotropy of our model imply that there is one
distinct equation for each of the four classes of edge in the lattice.

The Regge equation associated with an individual timelike edge $m_i^2$
involves six timelike lattice faces, and has the form
\[
0=2\, \epsilon^{xt}_i\, \pderiv{A ^{xt}_i}{m ^2_i} +
  2\, \epsilon^{yt}_i\, \pderiv{A ^{yt}_i}{m ^2_i} + 
  2\, \epsilon^{zt}_i\, \pderiv{A ^{zt}_i}{m ^2_i},
\]
which is the discrete counterpart of (\ref{eqn:kasnert}), the Einstein
field equation $G_{tt}=0$.  Similarly, the Regge equation which
corresponds to the single spatial edge $u_i$ involves six lattice
faces: two spacelike faces with area $u_iv_i$, two with area $u_i
w_i$, plus the timelike faces formed by evolving $u_i$ both forwards
and backwards in time.  The vacuum Regge equation is
\[
0=2\, \epsilon^{xy}_i\, \pderiv{A^{xy}_i}{u^2_i} 
  + 2\, \epsilon^{xz}_i\, \pderiv{A^{xz}_i}{u^2_i} 
  + \epsilon^{xt}_i\, \pderiv {A^{xt}_i}{u_i^2}
  + \epsilon^{xt}_{i-1}\, \pderiv {A^{xt}_{i-1}}{u_i^2}
\]
with similar expressions for the equations arising from the spacelike
edges $v_i$ and $w_i$.

Using the geometric information collected in section \ref{sec:lattice}
we obtain a single Regge equation for each class of edge on the $i$th
hypersurface, namely
\begin{eqnarray}
%0& = &
\fl \qquad
0=  -\frac{(w_i+w_{i+1})\, \epsilon ^{{zt}}_i}{4 \sqrt{\Delta w_i^2-4m_i^2}}
  -\frac{(v_i+v_{i+1})\, \epsilon ^{{yt}}_i}{4 \sqrt{\Delta v_i^2-4m_i^2}}
  -\frac{(u_i+u_{i+1})\, \epsilon ^{{xt}}_i}{4 \sqrt{\Delta u_i^2-4m_i^2}} 
  \label{eqn:regget} \\
%0 & = &
\fl\qquad  
0=v_i \epsilon ^{xy}_i +w_i   \epsilon ^{xz}_i
 + \frac{ -2 m_{i-1}^2 + u_i \Delta u_{i-1} } 
  {2  \sqrt{\Delta u_{i-1}-4m_{i-1}^2}}\, \epsilon ^{{xt}}_{i-1}
  + \frac{ - 2m_i^2 - u_i \Delta u_i }
  {2 \sqrt{\Delta u_i^2-4m_i^2}}\, \epsilon ^{xt}_i   
  \label{eqn:reggex} \\
%0 & = &
\fl\qquad  
0=  u_i \epsilon ^{xy}_i + w_i   \epsilon ^{yz}_i
  + \frac{ -2 m_{i-1}^2 + v_i \Delta v_{i-1} } {2  \sqrt{\Delta v_{i-1}-4m_{i-1}^2}}\, \epsilon ^{{yt}}_{i-1}
  + \frac{ - 2m_i^2 - v_i \Delta v_i } {2 \sqrt{\Delta v_i^2-4m_i^2}}\, \epsilon ^{yt}_i
  \label{eqn:reggey} \\
%0& = &
\fl\qquad
0=  u_i \epsilon ^{xz}_i  +v_i   \epsilon ^{yz}_i
 + \frac{ -2 m_{i-1}^2 + w_i \Delta w_{i-1} } 
 {2  \sqrt{\Delta w_{i-1}-4m_{i-1}^2}}\, \epsilon ^{{zt}}_{i-1}
  + \frac{ - 2m_i^2 - w_i \Delta w_i }
  {2 \sqrt{\Delta w_i^2-4m_i^2}}\, \epsilon ^{zt}_i 
  \label{eqn:reggez}
\end{eqnarray}
which correspond to the lattice edges $m_j^2$, $u_j$, $v_j$ and $w_j$,
respectively.

The structure of the Regge equations
(\ref{eqn:regget})-(\ref{eqn:reggez}) is worth considering.  Equation
(\ref{eqn:regget}), associated with the timelike edge $m_i^2$,
involves edges on, and between, two neighbouring hypersurfaces,
whereas (\ref{eqn:reggex})-(\ref{eqn:reggez}) involve information on
and between three consecutive spatial hypersurfaces.  Thus
(\ref{eqn:regget}) is a first-order constraint, while
(\ref{eqn:reggex})-(\ref{eqn:reggez}) are second-order difference
equations.

This contrasts sharply with the equations derived by Lewis
\cite{lewis82}, who neglected the curvature associated with spacelike
hinges, and was thus unable to derive the spatial equations
(\ref{eqn:reggex})-(\ref{eqn:reggez}).  After correctly obtaining
(\ref{eqn:regget}), Lewis found that he could only make sense of the
truncated spatial equations by considering a careful average. This
averaging resulted, once again, in the timelike equation
(\ref{eqn:regget}).  Without the spatial curvature terms
$\epsilon^{yz}_i$, $\epsilon^{xz}_i$ and $\epsilon^{xy}_i$ Lewis was
unable to build the second-order Regge equations 
(\ref{eqn:reggex})-(\ref{eqn:reggez}).

\begin{figure}[t]
  \centering
  \subfigure[Evolution of edge lengths]{
    \includegraphics[width=0.42\textwidth]{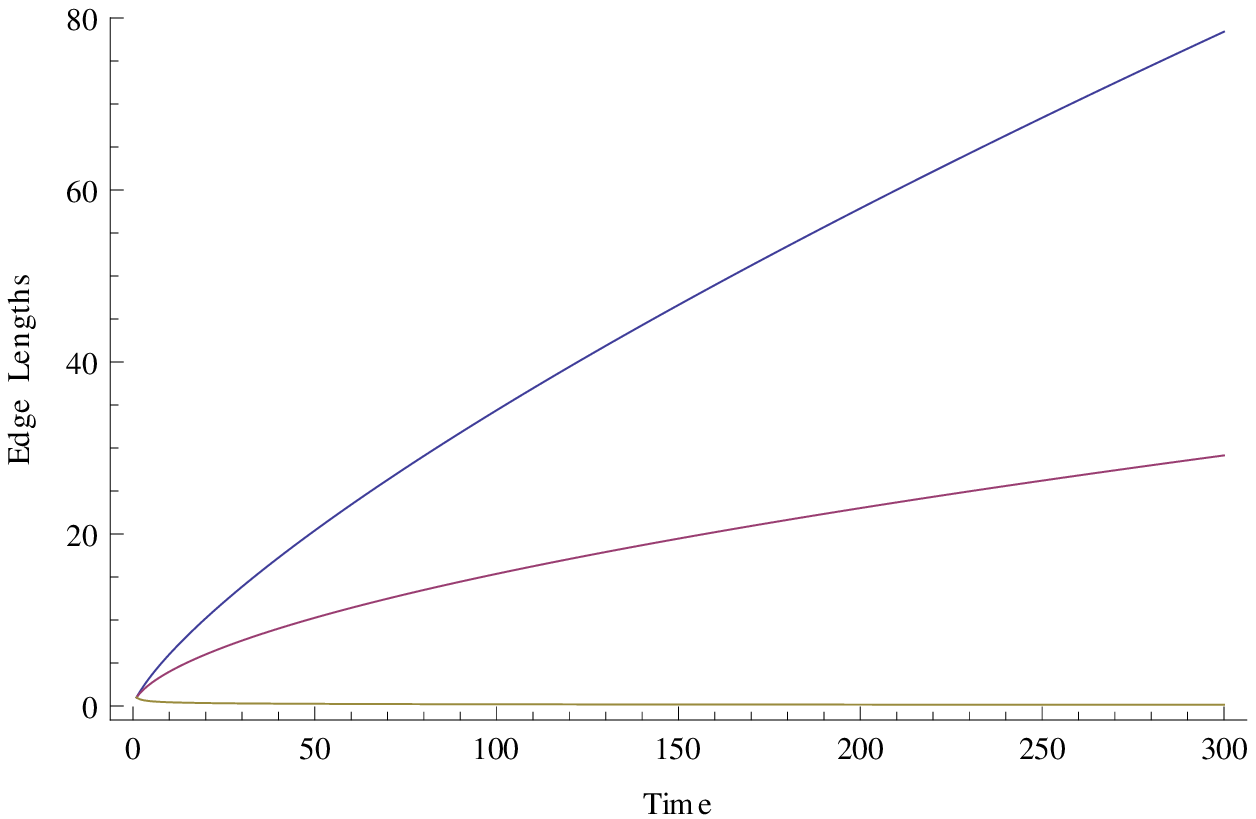}
    \label{fig:subfig2.1}  }
  \subfigure[Fractional error in edge lengths]{
    \includegraphics[width=0.45\textwidth]{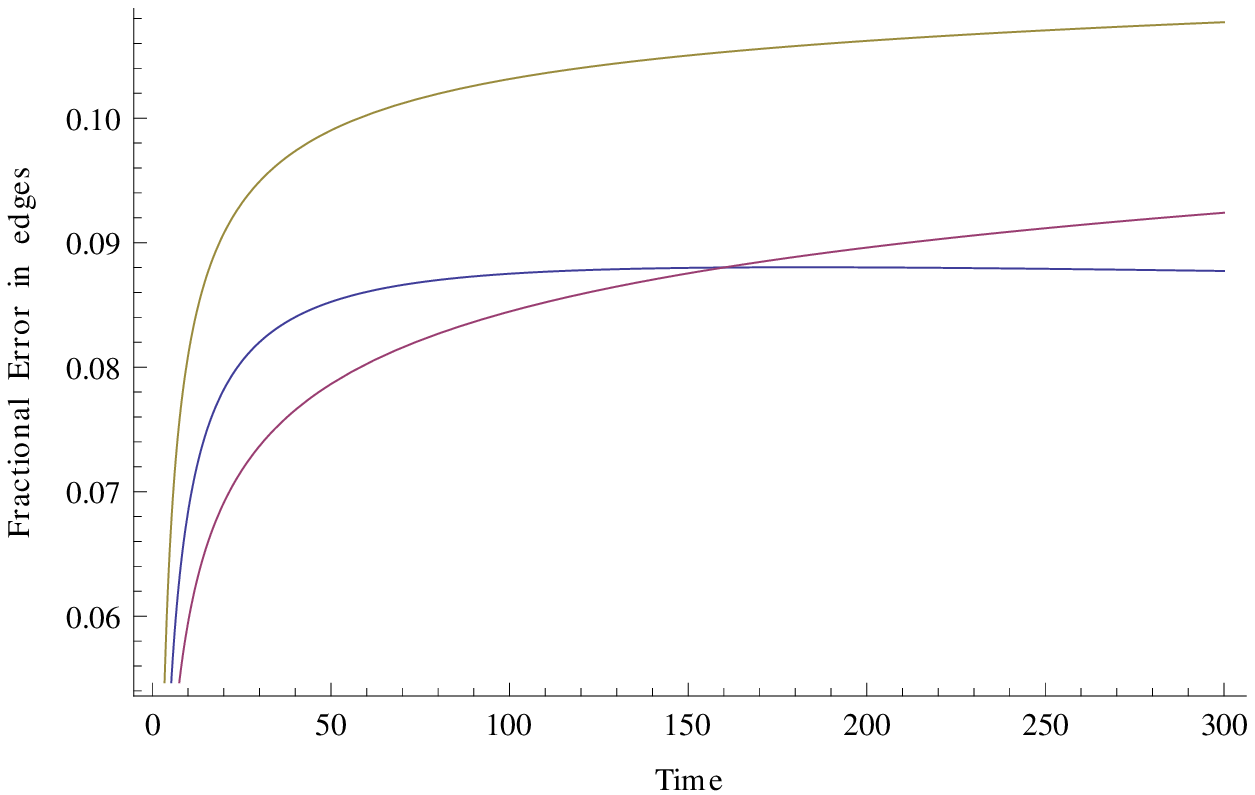}
    \label{fig:subfig2.3}  }
  \subfigure[Time evolution of the constraint $R_t$]{
    \includegraphics[width=0.48\textwidth]{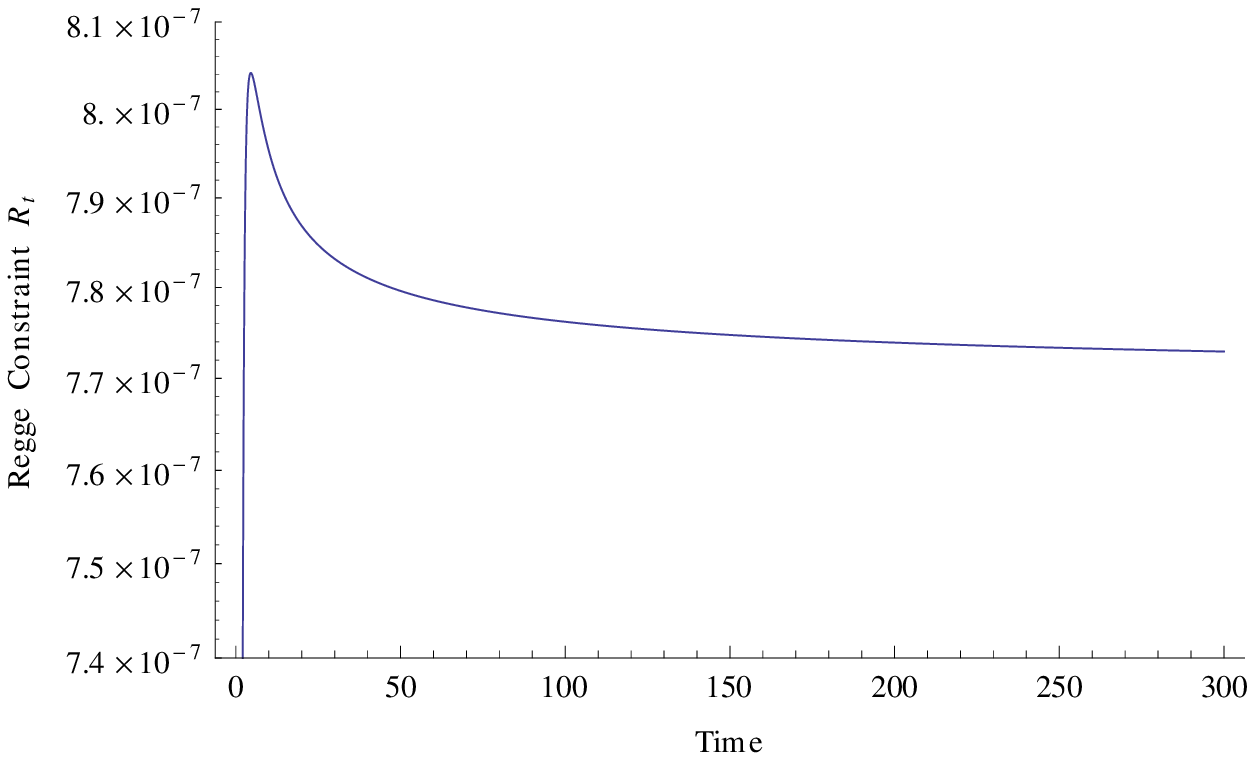}
    \label{fig:subfig2.2}  }
  \subfigure[Convergence of the mean value of $R_t$]{
    \includegraphics[width=0.45\textwidth]{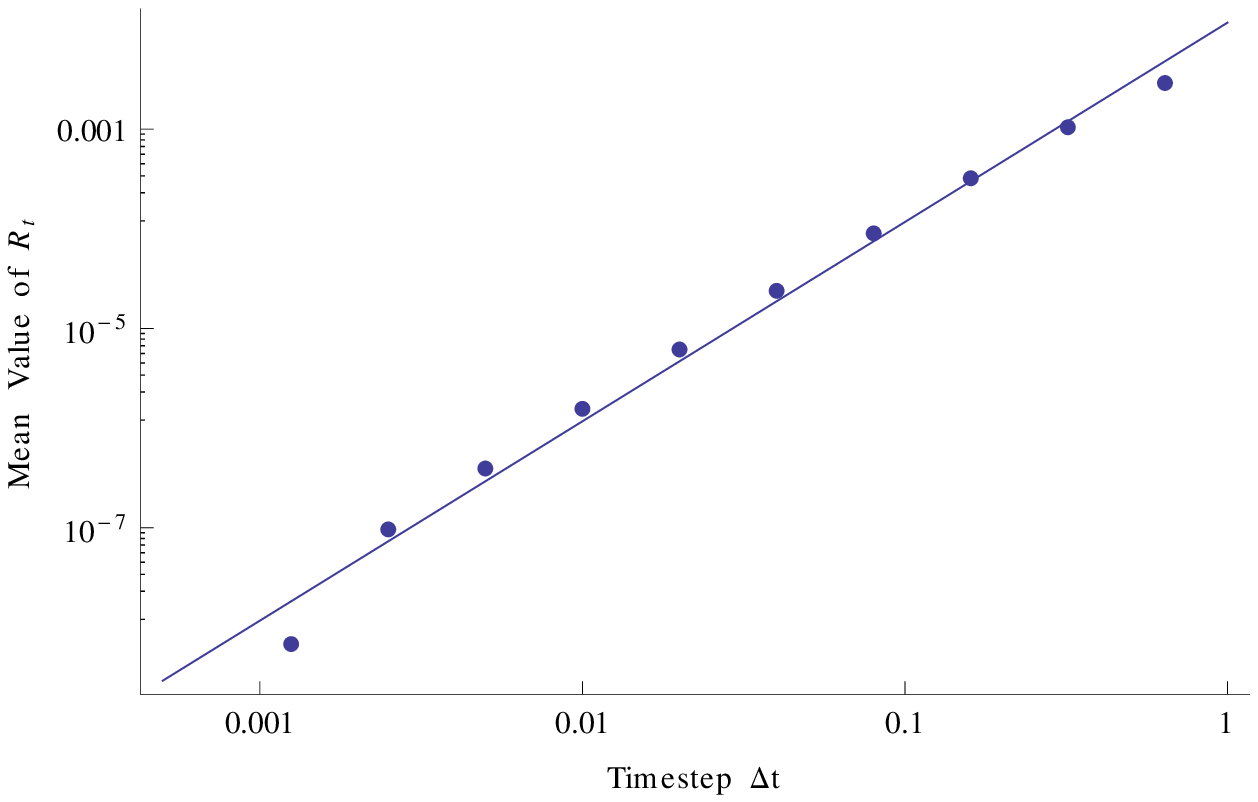}
    \label{fig:subfig2.4}  }
  \caption[]{Discrete Regge solutions with $p_1 =0.75$ and $\Delta
    t=0.01$.}
  \label{fig:discrete_solutions}
\end{figure}

Before examining the continuum limit of the Regge model we consider
solutions of the discrete equations.  Initial data is constructed at
$t_0=1$ to match the continuum Kasner solution as far as possible.
Taking the exact initial data to be $x(1)=y(1)=z(1)=1$ and $\dot
x(1)=p_1$, $\dot y(1)=p_2$, and $\dot z(1)=p_3$, we mimic the
properties of the exact solution in the lattice by setting
$u_0=v_0=v_0=1$ and
\[
\fl \qquad
 u_1 = u_0 + ( p_1+\alpha)\,\Delta t, \qquad
v_1 = v_0 + ( p_2+\alpha)\,\Delta t, \qquad 
w_1 = w_0 + ( p_3-2\alpha)\,\Delta t
\]
where $\alpha$ is an unknown parameter.  This form is chosen to
maintain the continuum condition $\dot x(1)+ \dot y(1)+ \dot z(1)=1$
to first order, which is physically equivalent to using the degrees of
freedom in the initial data to generate linear expansion in the volume
element.  With these initial data the Regge constraint
(\ref{eqn:regget}) is solved for the single parameter $\alpha$.

A typical solution of the discrete Regge equations is shown in figure
\ref{fig:discrete_solutions} for the case $p_1=0.75$ and $\Delta
t=0.01$.  The solution to the initial value problem in this case is
$\alpha=0.0214379$, which represents a roughly $3\%$ change in the
initial rate of change of $u_i$ compared to the exact estimate.  The
evolution of the initial data is shown in figure
\ref{fig:discrete_solutions}a, while \ref{fig:discrete_solutions}b
shows the evolution of the fractional error in the Regge solutions
compared with their exact counterparts.  The fractional error in all
edges remains in the $5\%-10\%$ range, and can be shown to shrink as
the time step is reduced. 

The residual error in the first-order Regge constraint
(\ref{eqn:regget}) is defined as
\begin{equation}
  R_t= \frac{(w_i+w_{i+1})\, \epsilon ^{{zt}}_i}{4 \sqrt{\Delta w_i^2-4m_i^2}}
  +\frac{(v_i+v_{i+1})\, \epsilon ^{{yt}}_i}{4 \sqrt{\Delta v_i^2-4m_i^2}}
  +\frac{(u_i+u_{i+1})\, \epsilon ^{{xt}}_i}{4 \sqrt{\Delta u_i^2-4m_i^2}},
  \label{eqn:reggeresidual} \\
\end{equation}
and is a measure of the consistency amongst the Regge
equations. Figure \ref{fig:discrete_solutions}c shows $R_t$ as a
function of time with $\Delta t=0.01$, and clearly the residual $R_t$
remains small throughout the evolution.  We repeated this process with
different $\Delta t$ to estimate the rate at which $R_t$ reduces as
$\Delta t$ tends to zero.  Figure \ref{fig:discrete_solutions}d shows
second-order convergence in the mean value of $R_t$ over $1<t<100$ as
the timestep is reduced.  We explore the issue of convergence in more
detail in the following sections.

% ************************

\section{The continuous-time Regge model}
\label{sec:reggecont}

Many of the early applications of Regge calculus to highly symmetric
spacetimes considered the differential equations that arise in the
limit of continuous time and discrete space
\cite{lewis82,collins73,brewin87}.  In this section we derive the
continuous time Regge equations and compare them with the results of
Lewis \cite{lewis82} before considering numerical solutions.

The continuous-time Regge model is developed from the discrete
equations in section \ref{sec:regge} in the limit of small $\Delta t$,
with the assumption that spatial edges in the lattice approach
continuous functions of time.  For example, the spacelike edge $u_i$
is viewed as the value of a continuous function $u(t)$ evaluated at
$t=t_i$.  A power series expansion then relates the edge lengths on
neighbouring surfaces,
\begin{equation}
  \label{eqn:series}
  u_{i+1} = u(t_i) + \pderiv{u}{t} \Delta t 
  + \frac{1}{2} \pderiv{^2u}{t^2} \Delta t^2 
  + \frac{1}{6} \pderiv{^3u}{t^3} \Delta t^3  
  + {\Or}(\Delta t^4) ,
\end{equation}
where all derivatives are evaluated at $t=t_i$.  Similar expressions
hold for the edges $u_{i-1}, v_{i+1}$ etc.  The series expansion for the
timelike edges $m_i^2$ is obtained from (\ref{eqn:timelike_edge}).

In the continuum limit the deficit angle about the spacelike area $A^{xy}_i$ is
\begin{eqnarray*}
  \epsilon_{xy}(t) &=& \frac{4 \ddot u}{4-\dot u^2}\,\Delta t + \Or(\Delta t^3), %%\\
%  \epsilon_{xz} &=& \frac{4 \ddot v}{4-\dot v^2}\,\Delta t + \Or(\Delta t^3),\\
%  \epsilon_{yz} &= &\frac{4 \ddot w}{4-\dot w^2}\,\Delta t + \Or(\Delta t^3),
\end{eqnarray*}
and the deficit angle about the timelike face formed by the evolution
of $u_i$ is
\begin{eqnarray*}
  \label{eqn:ctsdeficits2}
  \epsilon_{xt}(t) &=& 2\pi - 4 \cos^{-1}
  \left(  \frac{\dot{v}\, \dot{w}}
    {\sqrt{(4 - \dot{v}^2) (4 - \dot{w}^2)}} \right) + \Or(\Delta t),  % \\
%  \epsilon_{yt}(t) &=& 2\pi - 4 \cos^{-1}
%  \left(  \frac{\dot{u}\, \dot{w}}
%    {\sqrt{(4 - \dot{u}^2) (4 - \dot{w}^2)}} \right) + \Or(\Delta t)\\
%  \epsilon_{zt}(t) &=& 2\pi - 4 \cos^{-1}
%  \left(  \frac{\dot{u}\, \dot{v}}
%    {\sqrt{(4 - \dot{u}^2) (4 - \dot{v}^2)}} \right)+ \Or(\Delta t).
\end{eqnarray*}
with similar expressions for the remaining faces. 

Using these expansions, the Regge equations
(\ref{eqn:regget})-(\ref{eqn:reggez}) are, to leading order,
\begin{eqnarray}
\fl \quad
0=-\left[ 
  \frac{ u\, \epsilon_{xt}^0}{\sqrt{4-\dot v^2-\dot w^2}}
  +\frac{ v\, \epsilon_{yt}^0}{\sqrt{4-\dot u^2-\dot w^2}}
  +\frac{ w\, \epsilon_{zt}^0}{\sqrt{4-\dot u^2-\dot v^2}}\, \right] + \Or(\Delta t) 
  \label{eqn:reggectst}\\
%%% x equation
\fl \quad
0=\left[\, \frac{4 w \ddot v}{4-\dot v^2} 
  +  \frac{4 v \ddot w}{4-\dot w^2} 
  - \frac{  4 u \dot u }{4-\dot v^2-\dot w^2}
  \left(  \frac{ \ddot v \dot w }{4-\dot v^2}
    + \frac{ \dot v \ddot w }{4-\dot w^2} \right) \right.
  \label{eqn:reggectsx}\\
  \left. \quad +\, \frac{\epsilon^0_{xt}\,  \left(4-\dot u^2-\dot v^2-\dot w^2-u\ddot u\right)}
  {2 \sqrt{4-\dot v^2-\dot w^2}}
  - \frac{\epsilon^0_{xt}\,  u \dot u \left(\dot v \ddot v+ \dot w \ddot w\right)}
  {2 \left(4-\dot v^2-\dot w^2\right)^{3/2}} \,
  \, \right] \Delta t + \Or(\Delta t^3)  \nonumber \\
%%% y equation
\fl
\quad 0=  \left[\,
  \frac{4 w \ddot u}{4-\dot u^2} + \frac{4 u\ddot w}{4-\dot w^2}
  - \frac{4 v \dot v}{4-\dot u^2-\dot w^2}
  \left( \frac{\ddot u \dot w}{4-\dot u^2}
    + \frac{ \dot u \ddot w} {4-\dot w^2} \right)\right.
  \label{eqn:reggectsy}\\
  \left. \quad +\, \frac{\epsilon^0_{yt} \left(4-\dot u^2-\dot v^2-\dot w^2-v\ddot v\right)}
  {2 \sqrt{4-\dot u^2-\dot w^2}}
  -\frac{\epsilon^0_{yt}  v \dot v \left(\dot u\ddot u+\dot w\ddot w\right)}
  {2 \left(4-\dot u^2-\dot w^2\right)^{3/2}}\,
  \, \right]    \Delta t + \Or(\Delta t^3) \nonumber \\
%%% z equation
\fl
\quad 0=\left[\, \frac{4 v \ddot u}{4-\dot u^2} + \frac{4 u \ddot v}{4-\dot v^2}
  - \frac{4 w\dot w}{4-\dot u^2-\dot v^2}
  \left( 
    \frac{\ddot u\dot v}  {4-\dot u^2}
    + \frac{ \dot u \ddot v}{4-\dot v^2}
  \right) \right.  \label{eqn:reggectsz} \\
  \left.   \quad +\,
  \frac{\epsilon^0_{zt} \left(4-\dot u^2-\dot v^2-\dot w^2-w\ddot w\right)}
  {2 \sqrt{4-\dot u^2-\dot v^2}}
  -\frac{\epsilon^0_{zt}\, w \dot w \left( \dot u\ddot u+ \dot v\ddot v\right)}
  {2 \left(4-\dot u^2-\dot v^2\right)^{3/2}}\,
  \, \right]    \Delta t+ \Or(\Delta t^3) \nonumber 
\end{eqnarray}
where $\epsilon^0_{xt}, \epsilon^0_{yt}$ and $\epsilon^0_{zt} $ denote
the zeroth order terms in the deficit angle expansions.  The leading
terms in (\ref{eqn:reggectst})-(\ref{eqn:reggectsz}) are the
continuous-time Regge equations, a set of non-linear differential
equations.  Equation (\ref{eqn:reggectst}) is a first-order
differential equation, while the remainder are second-order.

\begin{figure}[t]
  \centering
  \subfigure[Evolution of edge lengths]{
    \includegraphics[width=0.45\textwidth]{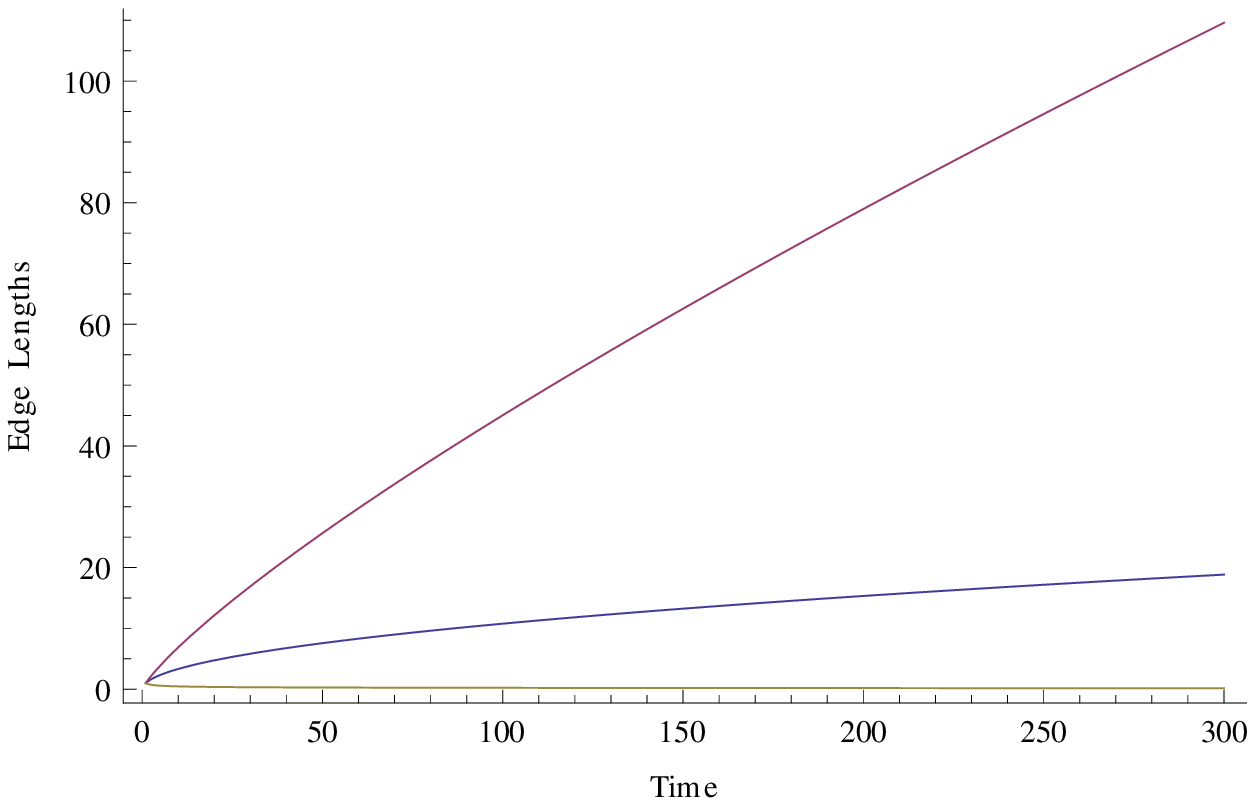}
    \label{fig:subfig3.1}  }
  \subfigure[Fractional error in edge lengths]{
    \includegraphics[width=0.45\textwidth]{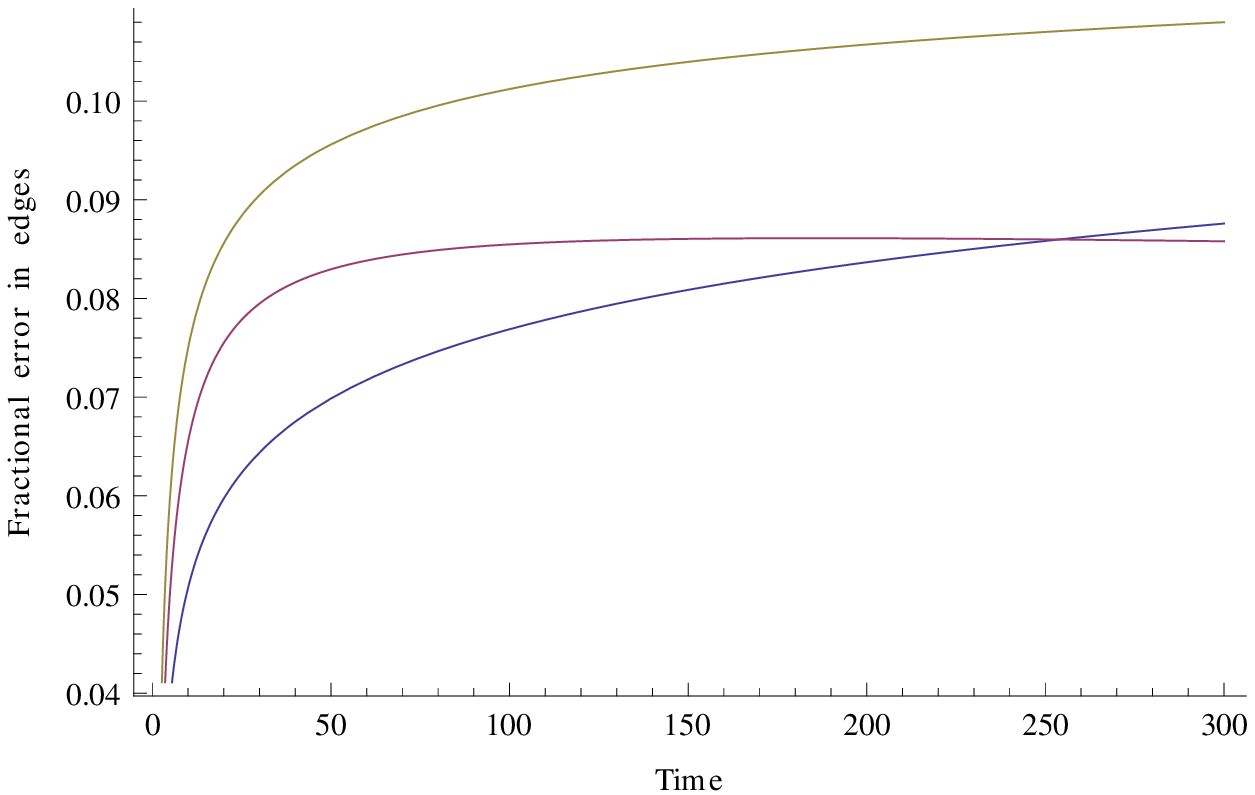}
    \label{fig:subfig3.2}  }
  \subfigure[Fractional error in prism volume]{
    \includegraphics[width=0.44\textwidth]{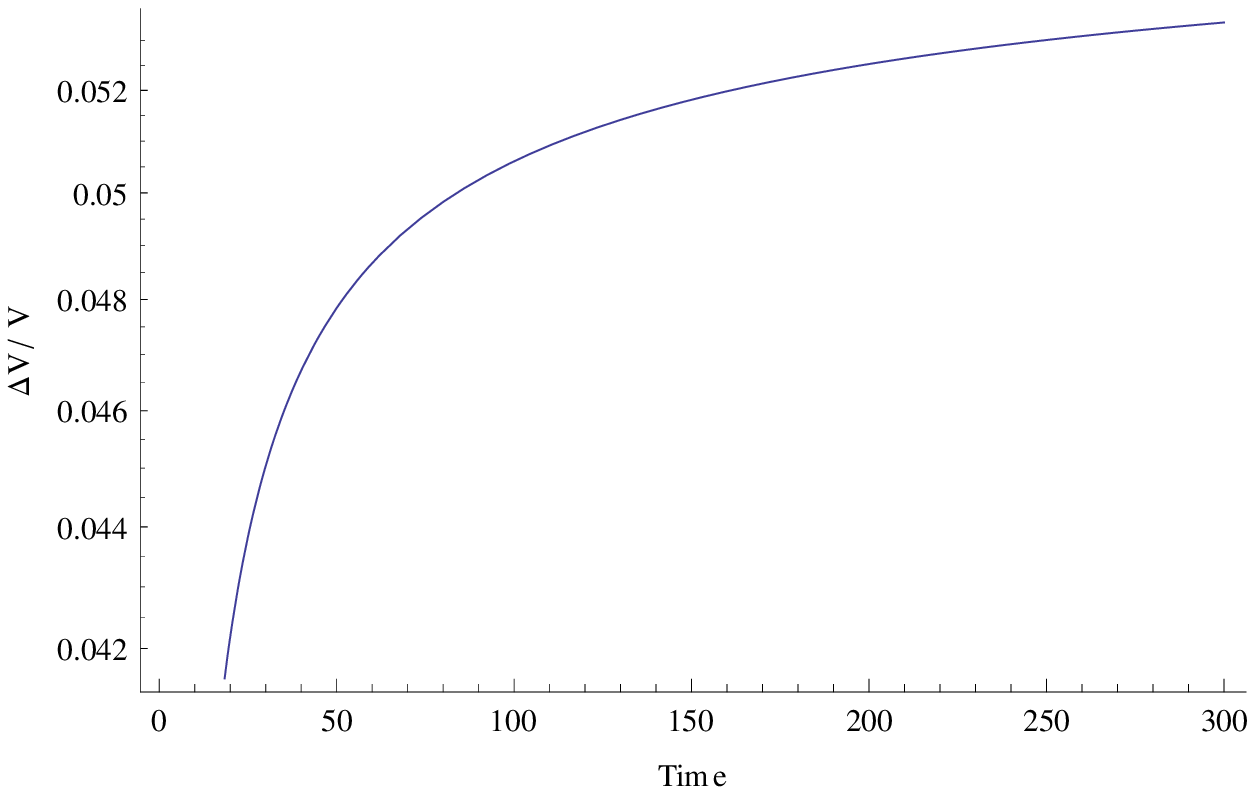}
    \label{fig:subfig3.3} } \subfigure[Residual in the constraint
  (\ref{eqn:reggectst})]{
    \includegraphics[width=0.47\textwidth]{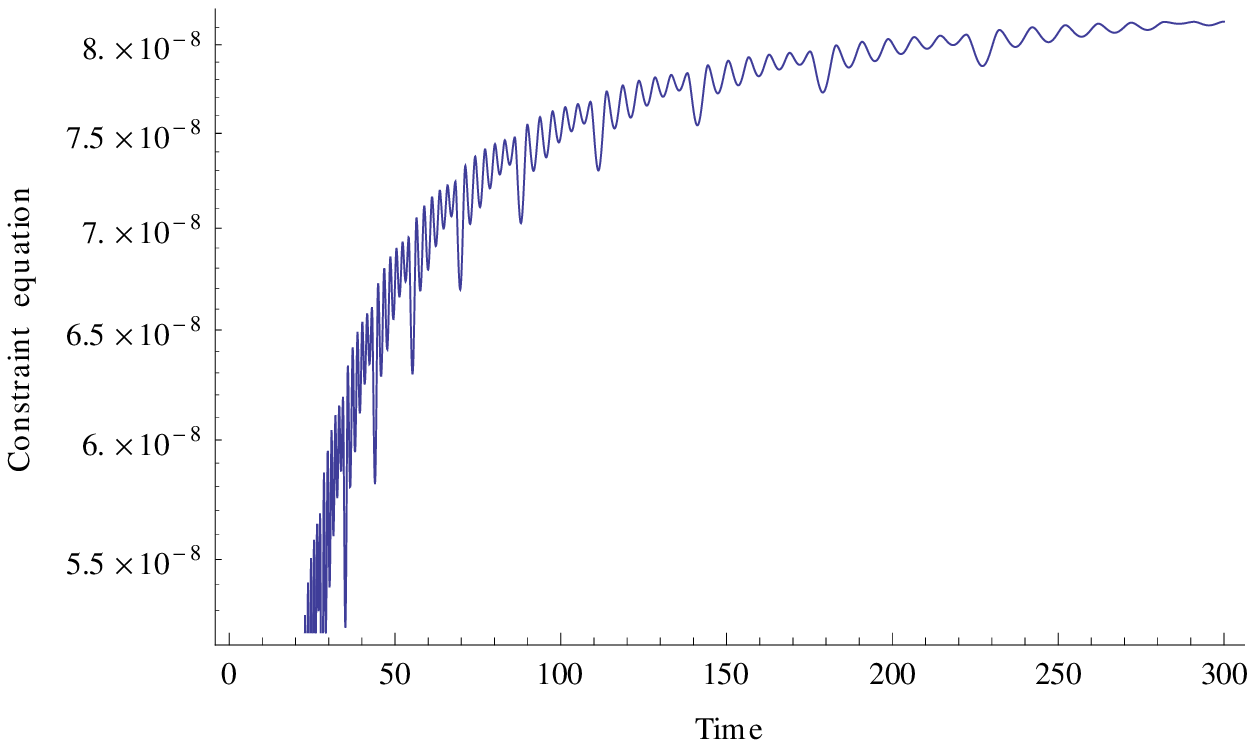}
    \label{fig:subfig3.4}  }
  \caption[]{Solution of the continuous-time Regge differential
    equations with $p_1 =0.5$.}
  \label{fig:cts_solutions}
\end{figure}

The continuous-time Regge differential equations were solved
numerically using {\sl Mathematica}.  Initial conditions were applied
at $t_0=1$, and the initial data was chosen to ensure that the
expansion of lattice volume elements is initially linear to match the
exact solution in section \ref{sec:kasner}.  Once again introducing a
parameter $\alpha$, we set
\begin{eqnarray*}
  u(1)=v(1)=w(1)=1 \qquad \mbox{and} \\
  \dot u(1)=p_1+\alpha, \qquad  \dot v(1)=p_2+\alpha, \qquad \dot w(1)=p_3-2\alpha,
\end{eqnarray*}
and use the first-order Regge initial value equation
(\ref{eqn:reggectst}) to solve for $\alpha$. This mimics the exact
initial data, for which $\alpha=0$.  The second-order, non-linear
differential equations (\ref{eqn:reggectsx})-(\ref{eqn:reggectsz}) are
used to evolve the initial data forward in time.

Figure \ref{fig:cts_solutions} shows solutions to the continuous-time
Regge equations with $p_1=0.5$.  The solution to the initial value
problem is $\alpha=0.0204161$, which represents a small deviation
($\approx 4\%$ change in the initial value of $\dot u$) from the exact
initial data. As can be seen in the figure, the continuous-time Regge
solutions are very similar to the exact Einstein solution, with the
Regge edges deviating from the exact values by $5\%-7\%$.  In the next
section we extend the limiting process to the spatial edges, and
examine the difference between the exact and Regge equations more
carefully.

%%%%%%%%%%%%%%%%%%%%%%%%%%%%%%%%%%%%%%%%%%%%%%%%%%

\section{The Regge equations in the limit of continuous space and
  time}
\label{sec:fullcts}

In this section we explore the discrepancies between the discrete
Regge model and the Kasner spacetime by examining the truncation error
incurred when the Regge equations are viewed as approximations of the
Kasner-Einstein equations (\ref{eqn:kasnert})-(\ref{eqn:kasnerz}).  We
consider the continuum limit of the Regge equations
(\ref{eqn:regget})-(\ref{eqn:reggez}) as the spatial lattice is
refined in both space and time.

The continuous space and time limit of the Regge equations is obtained
from the temporal series expansions (\ref{eqn:series}) together with
the link between lattice edges and the global length scales given by
(\ref{eqn:ratios}).  Substituting these into the discrete Regge
equations (\ref{eqn:regget})-(\ref{eqn:reggez}) and simultaneously
increasing the number of prisms ($n \rightarrow \infty$) while
reducing the timestep ($\Delta t \rightarrow 0$) we obtain a series
expansion for the the discrete equations in the continuum limit.  The
refinement parameter $n$ and timestep $\Delta t$ are chosen so that
$m_j^2 <0$.

The series expansion of the temporal Regge equation (\ref{eqn:regget}) is
\begin{eqnarray}
  \label{eqn:reggecontt}
%  \lefteqn{ 
  \fl
  \qquad 0 =  \frac{-1}{2n^3}\left[  \left( x\deriv{y}{t}\deriv{z}{t}  + y \deriv{x}{t} \deriv{z}{t} +
      z \deriv{x}{t}\deriv{y}{t}\right) \right. \\
   \left.\quad +\, \frac{1}{2}
  \left( \ 3\, \deriv{x}{t}\deriv{y}{t}\deriv{z}{t} 
    + z\deriv{y}{t}\deriv{^2x}{t^2} 
    + y\deriv{z}{t}\deriv{^2x}{t}  
    + z\deriv{x}{t}\deriv{^2y}{t}  \right.
  \right. \nonumber \\
    \left. \left. \qquad\qquad   +\ x\deriv{z}{t}\deriv{^2y}{t} 
    + y\deriv{x}{t}\deriv{^2z}{t} 
    + x\deriv{y}{t}\deriv{^2z}{t} 
  \right)\, \Delta t
   + {\Or}(\Delta t^2, \frac{1}{n^2})\right],  \nonumber
\end{eqnarray}
which has truncation error that is first-order in the timestep $\Delta
t$ and second-order in the spatial discretization scale $1/n$.  The
spatial Regge equations (\ref{eqn:reggex})-(\ref{eqn:reggez}) are
\begin{eqnarray}
  \label{eqn:reggecontx}
  \frac{\Delta t}{n^2} \left\{ 
    \deriv{y}{t}\deriv{z}{t} 
    + y\deriv{^2z}{t^2} + z\deriv{^2y}{t^2}  
    + {\Or}( \Delta t^2, \frac{1}{n^2} ) \right\} =0 \\
  \label{eqn:reggeconty}
  \frac{\Delta t}{n^2} \left\{ 
    \deriv{x}{t}\deriv{z}{t} 
    + x\deriv{^2z}{t^2} + z\deriv{^2x}{t^2}  
    + {\Or}( \Delta t^2, \frac{1}{n^2} )\right\} =0 \\
  \label{eqn:reggecontz}
  \frac{\Delta t}{n^2} \left\{ 
    \deriv{x}{t}\deriv{y}{t} 
    + x\deriv{^2y}{t^2} + y\deriv{^2x}{t^2}  
    + {\Or}( \Delta t^2, \frac{1}{n^2} ) \right\}=0
\end{eqnarray}
to leading order in the continuum limit.  The truncation error in
these equations is second order in both the spatial and temporal
discretization scales.

The leading order terms in the continuous time and space Regge
equations (\ref{eqn:reggecontt})-(\ref{eqn:reggecontz}) are identical
to the Kasner-Einstein equations
(\ref{eqn:kasnert})-(\ref{eqn:kasnerz}), so we expect solutions of the
discrete lattice equations to approach the continuum solutions as
length scales in the lattice are reduced.  It is clear from the
preceding equations that the truncation error for the Regge
equations, when viewed as approximations to the Kasner-Einstein
equations, are second order in the spatial discretization scale $1/n$.
The truncation error is also second order in $\Delta t$ for the
spatial Regge equations (\ref{eqn:reggecontx})-(\ref{eqn:reggecontz}).

The truncation error in (\ref{eqn:reggecontt}) implies that the Regge
initial value equation (\ref{eqn:regget}) is only a first-order
approximation to its continuum counterpart.  This conflicts with the
calculations in section \ref{sec:regge}, where the truncation error in
the Regge constraint $R_t$ was found to converge to zero as the second
power of $\Delta t$ (see figure \ref{fig:discrete_solutions}d).  To
understand this contradiction, we rewrite the coefficients of $\Delta
t$ in the expansion (\ref{eqn:reggecontt}) as
\begin{eqnarray}
  \label{eqn:cont_kasner_Gtt2}
%  3\, \deriv{x}{t}\deriv{y}{t}\deriv{z}{t} 
%  + \deriv{y}{t} \left( z \deriv{^2x}{t^2} + x\deriv{^2z}{t} \right) 
%  + \deriv{z}{t} \left( y \deriv{^2x}{t^2} + x\deriv{^2y}{t} \right)
%  + \deriv{x}{t} \left( z \deriv{^2y}{t^2} + y\deriv{^2z}{t} \right)   \nonumber \\
\fl\quad  3\, \dot x \dot y \dot z 
  + \dot y \left( z \ddot x + x \ddot z \right) 
  + \dot z \left( y \ddot x + x \ddot y \right)
  + \dot x \left( z \ddot y + y \ddot z \right)   \nonumber \\
   =
  \dot x \left( \dot y \dot z + z \ddot y + y \ddot z \right)
  + \dot y \left( \dot x \dot z + z \ddot x + x \ddot z \right) 
  + \dot z \left( \dot y \dot z + y \ddot x + x \ddot y \right)   \nonumber \\
%    \deriv{x}{t} \left\{  \deriv{y}{t}\deriv{z}{t}  
%      + y\deriv{^2z}{t^2} + z\deriv{^2y}{t^2} \right\}
%    + \deriv{y}{t} \left\{  \deriv{x}{t}\deriv{z}{t} 
%      + x\deriv{^2z}{t^2} + z\deriv{^2x}{t^2} \right\} \nonumber \\
%    \qquad +  \deriv{z}{t} \left\{  \deriv{x}{t}\deriv{y}{t} 
%      + x\deriv{^2y}{t^2} + y\deriv{^2x}{t^2} \right\} 
  = 0 + {\Or}(\Delta t^2, \frac{1}{n^2} ), \nonumber 
\end{eqnarray}
where the final equality follows from substitution of
(\ref{eqn:reggecontx})-(\ref{eqn:reggecontz}).  Thus the truncation
error in the Regge constraint (\ref{eqn:reggecontt}) is formally of
order $\Delta t$, but the coefficient of that term in the expansion is
a linear combination of the spatial Kasner-Einstein equations.  This
should be zero to leading order for any solution of the spatial Regge
equations.

To clarify this argument, consider again the simulation in section
\ref{sec:regge}.  Once the initial data is set, the numerical
evolution is achieved by the repeated solution of the spatial Regge
equations (\ref{eqn:reggex})-(\ref{eqn:reggez}).  These are shown
above to be second order accurate approximations of the Einstein
equations in both space and time.  We expect from
(\ref{eqn:reggecontt}) that the leading order error in the Regge
constraint is first order in $\Delta t$.  However, the coefficient of
that error term is a linear combination of the Regge equations we are
solving (to leading order), and thus the {\sl coefficient} of $\Delta
t$ is itself zero to second order in $\Delta t$.  Thus the {\sl
  effective} leading order truncation in the Regge constraint equation
(\ref{eqn:reggecontt}) is of second order in both space and time.
This is consistent with the numerical experiments in section
\ref{sec:regge}.

\section{Discussion}

In the preceding sections we re-examined one of the most highly
symmetric applications of Regge calculus to be found in the
literature. The primary goal of this study was to examine the
convergence properties of Regge calculus, and we have shown that for
the discrete Kasner spacetime the equations of Regge calculus reduce
identically to the corresponding Einstein equations in the continuum
limit.

The discrete lattice used by Lewis, outlined above, was specifically
designed to guarantee a one-to-one correspondence between the degrees
of freedom in the Regge lattice and the metric components in the
continuum solution \cite{lewis82}.  Despite this, Lewis needed to
average the Regge evolution equations in order to obtain consistency
in the continuum limit.  The averaging process was chosen to obtain
the first order Regge equation (\ref{eqn:regget}), but Lewis was still
unable to derive the remaining spatial equations
(\ref{eqn:reggex})-(\ref{eqn:reggez}).  We showed in section
\ref{sec:regge} that once all lattice curvature elements are included
in the calculations the Regge equations consist of one constraint and
three evolution equations.  In section \ref{sec:fullcts} we showed
that these lattice equations approach the full set of Kasner-Einstein
equations in the continuum limit.

It was also shown in section \ref{sec:fullcts} that the discrete Regge
equations are second order accurate approximations to the Einstein
equations for the Kasner cosmology in the limit of very fine
discretization.  This convergence rate is in agreement with many
previous numerical simulations, in particular the (3+1)-dimensional
Regge calculus models of the Kasner cosmology that utilized general
simplicial lattices \cite{gentle98,brewin01}.  Unlike the current
analysis, these simulations did not enforce the homogeneity and
anisotropy of the Kasner model throughout the evolution, yet displayed
second-order convergence to the continuum solution.  These numerical
simulations considered the convergence of solutions, rather than
equations, and so are complementary to our analysis.

In general applications of Regge calculus that utilize a simplicial
lattice there will be many more Regge equations (one per lattice edge)
than Einstein equations (10 per spacetime event).  The direct
comparison between individual Regge and continuum equations considered
in this paper would not be possible, or even desirable.  We expect
that an appropriate average of the Regge equations would still
correspond to the Einstein equations in the continuum limit
\cite{miller86,brewin00,cheeger84}, and several such averaging schemes
have been suggested. Brewin considered a finite-element integration of
the weak-field Einstein equations over a simplicial lattice (discussed
in \cite{brewin00}), and suggested that the vertex-based equivalent of
the vacuum Einstein equations are
\[ 
0= \sum_j \sum_{i(j)} (\Delta x^\mu \Delta x^\nu)_i\, \epsilon_j\,
\frac{\partial A_j}{\partial L^2_i},
\]
where $\Delta x^\mu_j$ is a vector oriented along edge $L_j$ in a
coordinate system based at vertex $v$. The outer summation is over all
triangles which meet at $v$, and the inner sum is over the edges on
each triangle.  These are essentially linear combinations of Regge
equations, together with boundary terms \cite{brewin00}.

Regardless of how one compares the continuum and discrete equations,
it is ultimately the solutions that are of interest.  The application
of Regge calculus to the Kasner cosmology discussed in this paper
demonstrates yet again that Regge calculus is a consistent
second-order accurate discretization of general relativity, providing
further support for the use of lattice gravity in numerical relativity
and discrete quantum gravity.

% --- REFERENCES ----------------------------------------------------

\end{document}